\documentclass[a4paper,11pt]{article}
\usepackage{pos}

\title{High statistical computation of the Landau gauge ghost-gluon vertex}

\author*[a,b]{Nuno Brito}
\author[a]{Orlando Oliveira}
\author[a]{Paulo J Silva}

\affiliation[a]{CFisUC, Department of Physics, University of Coimbra, Portugal}
\affiliation[a]{Centre for Mathematical Sciences, University of Plymouth, UK}

\emailAdd{nmrbrito2000@gmail.com}
\emailAdd{orlando@uc.pt}
\emailAdd{psilva@uc.pt}

\abstract{The lattice computation of the one-particle irreducible 
ghost-gluon Green function in the Landau gauge is revisited with a set of large gauge ensembles. The large statistical ensembles enable a precise determination of this Green function over a wide range of momenta, accessing its IR and UV properties with a control on the lattice effects.
}

\FullConference{The 41st International Symposium on Lattice Field Theory (LATTICE2024)\\
 28 July - 3 August 2024\\
Liverpool, UK\\}

\begin{document}
\maketitle

\section{Motivation and Definitions}

In QFT the primary quantities to be computed are the Green functions given by vacuum expectations values of the quantum fields. In particular, 
the one-particle irreducible Green functions (1PI) summarize the dynamics of the theory. 
For QCD and for two-point functions, i.e. for the propagators of the fundamental fields (quarks, gluons, ghosts), we have achieved a good understanding of 
the non-perturbative features of the 1PI over a wide range of momenta that includes the UV regime. Current efforts are  on 
the non-perturbative evaluation of the 1PI functions with more than two external legs. The quark-gluon vertex, the three gluon vertex, the four gluon vertex are
examples of such type of functions.
Herein, we report on preliminary lattice data, using the pure Yang-Mills SU(3) theory in the Landau gauge, for the ghost-gluon 1PI.
This Green function has the diagrammatic representation
\begin{figure}[h]
      \centering
	  \includegraphics[width=4cm]{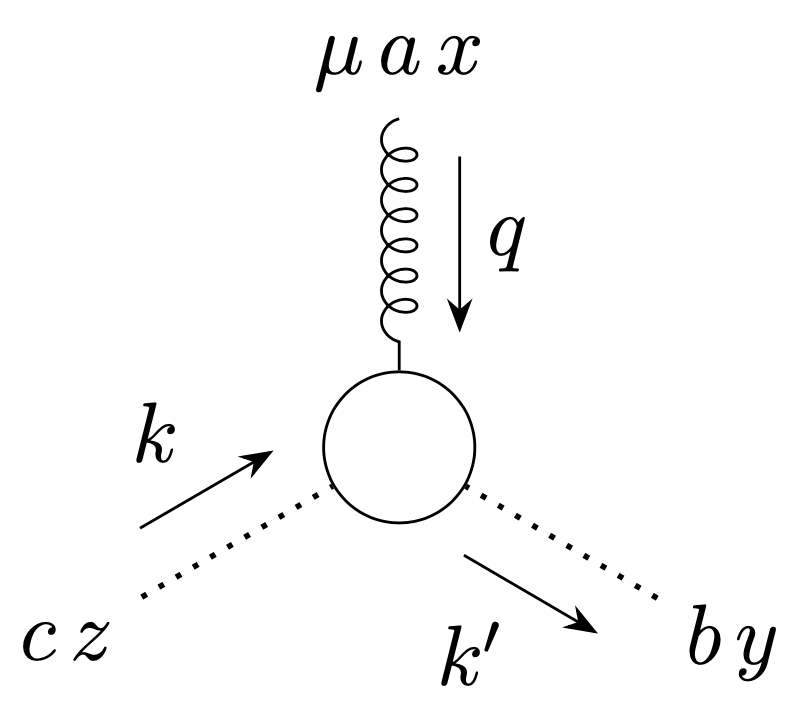}
\end{figure}

\noindent
with the corresponding Feynman rule being
\begin{equation}
    \Gamma^{abc}_\mu (q, \, k^\prime, \, k) = i \, g \, f_{abc} \bigg( k^\prime_\mu \, H_1(k^2, \, k^{\prime  ^2}, \, q^2 ) + q_\mu \, H_2(k^2, \, k^{\prime  ^2}, \, q^2 )\bigg) \ ,
    \label{vertex}
\end{equation}
where $g$ is the strong coupling constant and the functions $H_i$  are Lorentz scalar form factors. $\Gamma^{abc}_\mu$ is important for the QCD dynamics and, for example,
plays a major role in the determination of the ghost propagator properties and also of the gluon propagator.

\section{The Lattice Green Function}

On a lattice simulation the quantities that are directly evaluated  are the full Green functions $\mathcal{G}$. Then, for the case of the ghost-gluon vertex,
to access the form factors associated with the 1PI namely $H_1$ and $H_2$, see Eq. (\ref{vertex}), one needs to compute
\begin{figure}[h]
      \centering
	  \includegraphics[width=9cm]{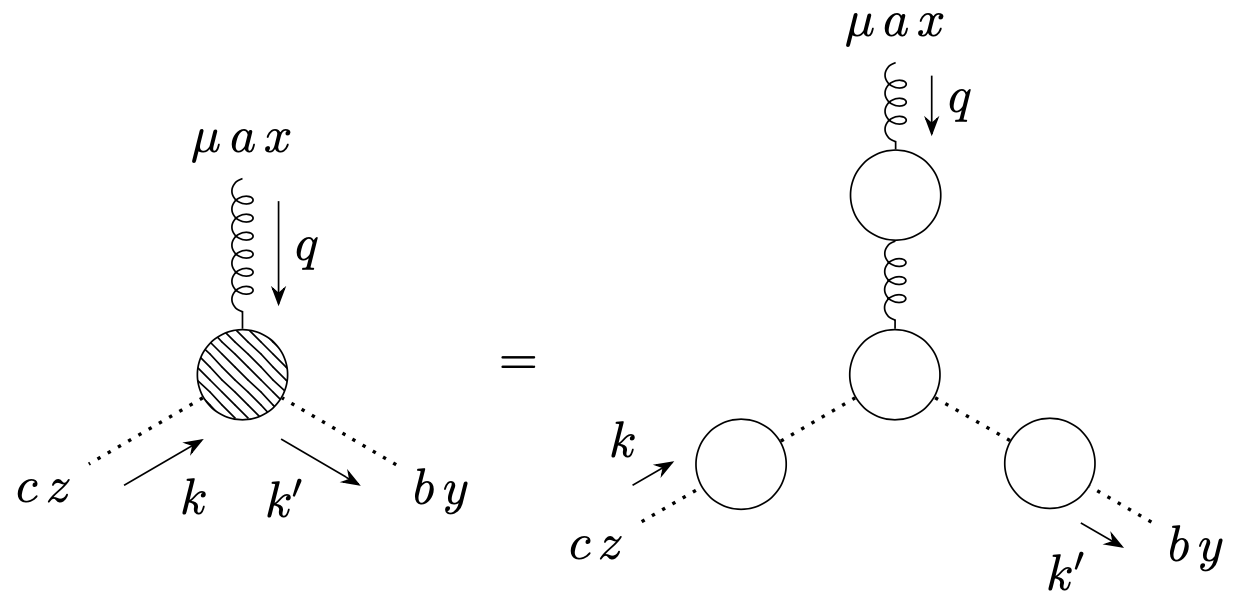} 
\end{figure}

\noindent
and, afterwards, disentangle the contribution of $H_1$ and $H_2$. 

In the Landau gauge,  the gluon propagator $D^{ab}_{\mu\nu}(k)$ is orthogonal to the momentum, i.e. its tensor structure is such that 
$k_\mu D^{ab}_{\mu\nu}(k) = k_\nu D^{ab}_{\mu\nu}(k) = 0$,  and is described by a single form factor. The orthogonality of the propagator implies that 
the contribution of the form factor $H_2$ to $\mathcal{G}$ vanishes. The full Green function is then determined uniquely by the scalar function $H_1$.

The perturbative solution of QCD provides an estimation for $\Gamma_\mu$ whose continuum and lattice regularized versions differ. 
Indeed, the tree level Feynman rule for the lattice regularized QCD is given by
\begin{displaymath}
    \Gamma^{(Lat) \, abc} _\mu = i \, g \, f_{abc} \, k^\prime_\mu \,  \cos \left( \frac{a \, k_\mu}{2}  \right)
\end{displaymath}
with
\begin{displaymath}
   k_\mu = \frac{ 2 \, \pi}{a \, N_\mu} \, n_\nu \ , \qquad n_\mu = 0, \dots, N_\mu/2 \ ,
\end{displaymath}
 $a$ being the lattice spacing, $N_\mu$ is the number of lattice points in direction $\mu$ and
\begin{equation}
k^\prime_\mu = (2/a) \, \sin ( a \, k_\mu / 2)
\end{equation}
is the improved momenta that reproduces the naive lattice momenta $k_\mu$ up to corrections of $\mathcal{O}(a^2)$. 
The tree level 1PI Green function $\Gamma^{(Lat)}_\mu$ matches the continuum rule in the limit of $a \rightarrow 0$ but it differs from its continuum expression
by a cosine-term and also by the momentum $k^\prime$ that differs from the naive lattice momentum.
 
 From the above definitions it follows that to estimate the form factor $H_1$ from the full Green function $\mathcal{G}$ the gluon and ghost propagators are required
 and have to be computed. Details and definitions used in the computation of the propagators can be found in ref. [4]. Once the propagators are known, the form factor 
 $H_1$ can be obtained  via the computation of the Lorentz-color contraction
\begin{equation}
   H_1  = \frac{ \Gamma_{(i)} \, \mathcal{G} }{ \Gamma_{(i)} \, D_{gl} \, D_{gh} \, D_{gh} \, \Gamma_{(i)} }  \ .
   \label{Contracao}
\end{equation}
In the last expression, the Lorentz and color indices as well as the momenta were omitted to simplify the notation.
$D_{gl}$ stands for the gluon propagator, $D_{gh}$ for the ghost propagator and the vertex
$\Gamma_{(i)}$ represents either $\Gamma^{(Lat)}$ of the continuum vertex $\Gamma^{(Cont)}$ (that differs
of $\Gamma^{(Lat)}$ by ignoring the $\cos$-term in the definition of $\Gamma^{(Lat)}$).
By using $\Gamma^{(Lat)}$ and $\Gamma^{(Cont)}$ we aim to estimate the effects coming from using a finite system to simulate QCD.

The preliminary results reported here use the following sets of gauge configurations ensembles
\begin{table}[h]
   \centering
   \begin{tabular}{ l @{\hspace{1.2cm}} c @{\hspace{1.2cm}} c} 
       \hline
      $\beta$  & $L$ & \# \\
       \hline 
      6.0     & 32     &  3000 \\
      6.0     & 48     &  2000 \\
       \hline 
   \end{tabular}
\end{table}

\noindent
that were generated using the Wilson action, rotated to the Landau gauge and used for studying the four gluon vertex [5]. 
The simulation is performed at a single $\beta$ value and, therefore,
we are not able to discuss the continuum limit, i.e. finite volume and lattice effects. 
However, the studies of the gluon propagator suggest that finite volume and lattice spacing effects are small
for the chosen $\beta$ value for such large lattices.
Computational simulations were performed with the Chroma  \cite{5} and PFFT \cite{6} libraries. 
For the conversion to physical units we take as lattice spacing $1/a = 1.943$ GeV 
(see [2] and references therein). Statistical errors were computed with the bootstrap method using a 67.5\% confidence level.

\section{Two and Three Point Functions}

\begin{figure}
	\centering
	\includegraphics[scale=0.5]{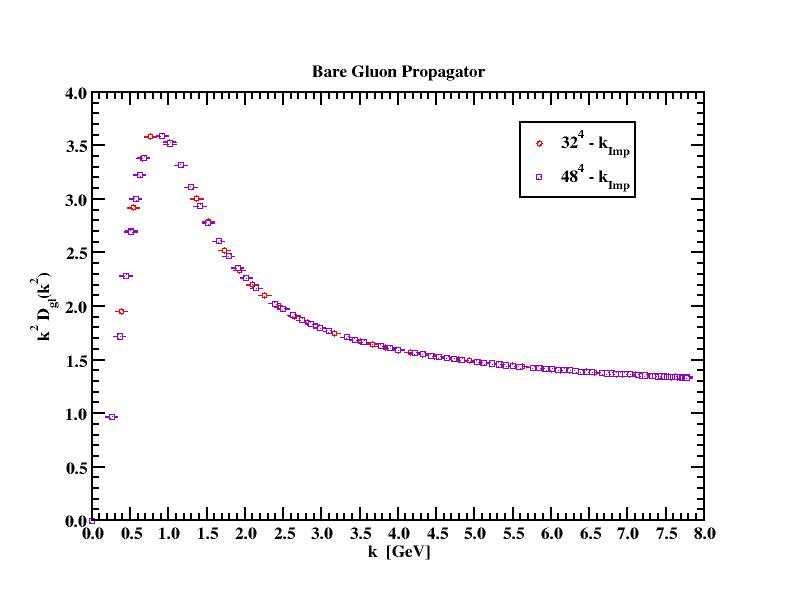}
	\caption{The bare gluon dressing function. \label{Gluon-Fig}}
\end{figure}

\begin{figure}
	\centering
	\includegraphics[scale=0.5]{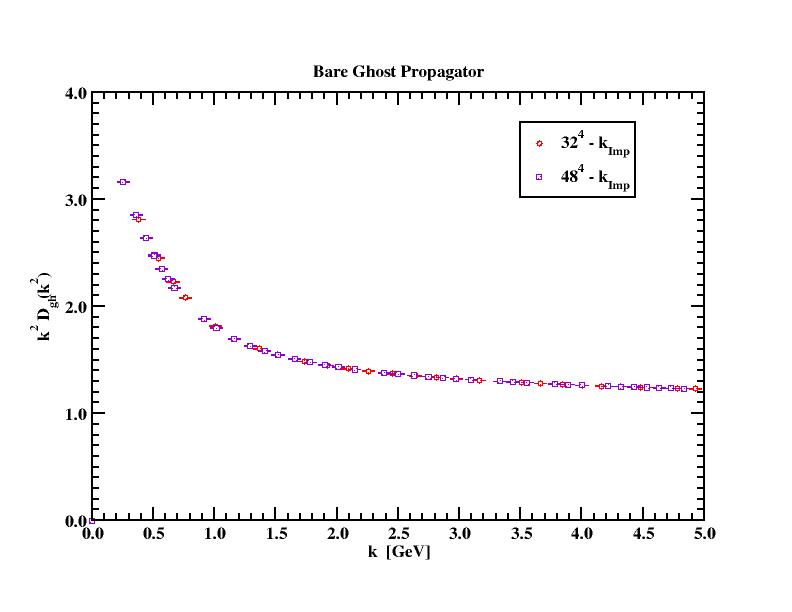}
        \caption{The bare ghost dressing function. \label{Ghost-Fig}}
\end{figure}

In the Landau gauge, the tensor structure of the gluon propagator is given by
\begin{equation}
   D^{ab}_{\mu\nu} (k) = \delta^{ab} \, \left( \delta_{\mu\nu} - \frac{k_\mu k_\nu}{k^2} \right) \, D_{gl}(k^2) \ ,
\end{equation}   
while the ghost propagator is described by
\begin{equation}
   D^{ab}  (k) = \delta^{ab} \, D_{gh}(k^2) \ .
\end{equation}   
Their dressing functions, as a function of the improved lattice momentum, are reported in Figs \ref{Gluon-Fig} and \ref{Ghost-Fig}, respectively. 
Note that the curves associated with $D_{gl}$ and $D_{gh}$ seem to be smooth functions of the momentum, while the statistical errors are tiny. 
The comparison of $D_{gl}$ for the two lattices shows a difference at zero momentum beyond the one standard deviation, that is not seen in the dressing function due 
to its definition. This difference has been reported before and is understood as a finite volume effect \cite{2,Oliveira:2012eh}.

\begin{figure}
	\centering
	\includegraphics[scale=0.4]{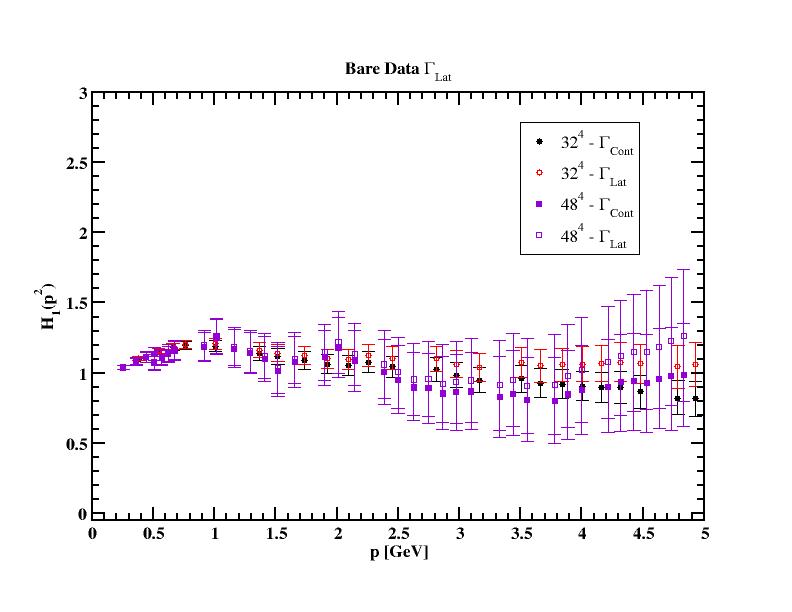}
	\caption{Preliminary bare lattice data for the form factor $H_1$. \label{H1-Fig}}
\end{figure}

So far, in our simulation, the form factor $H_1$ was computed only for the soft gluon limit that is defined by setting the gluon momentum to zero ($q = 0$). 
The three point function form factor is
reported in Fig. \ref{H1-Fig}. The bare lattice data shows a good agreement between the various simulations and for $\Gamma^{(Lat)}$ and $\Gamma^{(Cont)}$ for
momentum up to $\sim 2.5$ GeV. We recall the reader that the data is to be considered preliminary and we aim to achieve a good understanding of the finite volume 
and spacing effects in the near future.

\section{Summary and Conclusions}	

Let us summarise and conclude on the results given. Our data for the 1PI ghost-gluon vertex show good agreement over a large range of momenta and specially in the
infrared region.
Moreover, when compared with previous lattice calculations of the vertex \cite{6,7,8,9,10}, despite the large statistical erros for the SU(3) case, our data seems to be 
compatible with the older results. The form factor $H_1$ seems to be flat with the lattice data suggesting that it decreases slightly at low momenta.
The finite volume/spacing effects for momentum $\gtrsim 3$ GeV  need to be understood and will be investigated. It would be helpful also to increase the 
statistical precision of the calculation for this vertex.

\section*{Acknowledgments}

This work was supported by  FCT – Funda\c{c}\~ao para a Ci\^encia e a Tecnologia, I.P., under Projects Nos. UIDB/04564/2020 (\url{https://doi.org/10.54499/UIDB/04564/2020}),
UIDP/04564/2020 (\url{https://doi.org/10.54499/UIDP/04564/2020}) and CERN/FIS-PAR/0023/2021 (\url{https://doi.org/10.54499/CERN/FIS-PAR/0023/2021}).
N. B. acknowledges travel support provided by UKRI Science and Tecnhology Facilities (ST/X508676/1). 
P. J. S. acknowledges financial support from FCT contract CEECIND/00488/2017. 
(\url{https: //doi.org/10.54499/CEECIND/00488/2017/CP1460/CT0030}). 
The authors acknowledge the Laboratory for Advanced Computing at the University of Coimbra (\url{http://www.uc.pt/lca}) for providing access to the HPC resources.
Access to Navigator was partly supported by the FCT Advanced Computing Projects 2021.09759.CPCA (\url{https://doi.org/10.54499/2021.09759.CPCA}), 2022.15892.CPCA.A2, 2023.10947.CPCA.A2.



\begin{thebibliography}{99}
\bibitem{1}
P.~J.~Silva and O.~Oliveira, 
Nucl. Phys. B \textbf{690}, 177-198 (2004)

\bibitem{2}
A.~G.~Duarte, O.~Oliveira and P.~J.~Silva,
Phys. Rev. D \textbf{94}, no.1, 014502 (2016)

\bibitem{3} 
H.~Suman and K.~Schilling,
Phys. Lett. B \textbf{373}, 314-318 (1996)

\bibitem{4}
A.~C.~Aguilar \textit{et al.}
[arXiv:2404.06496 [hep-lat]].

\bibitem{5}
R. G. Edwards \textit{et al.},
Nucl. Phys. B Proc. Suppl. 140, 832 (2005). 

\bibitem{6}
M. Pippig, SIAM J. Sci. Comput. 35, C213 (2013).

\bibitem{Oliveira:2012eh}
O.~Oliveira and P.~J.~Silva,
Phys. Rev. D \textbf{86}, 114513 (2012)
[arXiv:1207.3029 [hep-lat]].

\bibitem{7}
M.~Cola\c{c}o, O.~Oliveira and P.~J.~Silva,
Phys. Rev. D \textbf{109}, no.7, 074502 (2024)

\bibitem{8} 
A.~Cucchieri, A.~Maas, and T.~Mendes, 
Phys. Rev. D 77, 094510 (2008).

\bibitem{9}
A.~Maas,
SciPost Phys. 8, 071 (2020)

\bibitem{10} 
E.~M.~Ilgenfritz \textit{et al.}
Braz. J. Phys. 37, 193 (2007)

\end{thebibliography}
\end{document}